\documentclass[aps,prl,twocolumn,titlepage,nofootinbib]{revtex4}
\usepackage{graphicx,physics}
\usepackage{color}
\usepackage{dcolumn}
\usepackage{latexsym}

\usepackage[normalem]{ulem}
\usepackage{hyperref,amssymb}
\usepackage{url}
\usepackage{color,soul}
\usepackage{graphicx}
\usepackage{verbatim}
\usepackage{multirow}
\usepackage{amsmath}
\newcommand{\beq}{\begin{eqnarray}}
\newcommand{\eeq}{\end{eqnarray}}
\usepackage{mathrsfs}
\usepackage{float,soul}
\usepackage[dvipsnames]{xcolor}
\usepackage{mathtools}
\usepackage{slashed}
\usepackage{physics}	
\usepackage{graphicx}   
\usepackage{epstopdf}
\usepackage{subfigure}  
\usepackage{hyperref}   
\usepackage{bbold}
\usepackage{wasysym}
\usepackage{feynmp}
\usepackage{natbib}
\hypersetup{colorlinks,
}

\usepackage[latin1]{inputenc}

\begin{document}

\title{\Large Real space solution of inhomogeneous elastic wave equation with localized vibration and flat dispersion relation}
\author{Da-Shan Jiang$^{1,2}$}
\email{cunyuanjiang@sjtu.edu.cn}
\address{$^1$ School of Physics and Astronomy, Shanghai Jiao Tong University, Shanghai 200240, China}
\address{$^2$ Wilczek Quantum Center, School of Physics and Astronomy, Shanghai Jiao Tong University, Shanghai 200240, China}

\begin{abstract}
The low frequency vibrational anomaly known as Boson peak (BP) have been studied extensively in various disordered systems, however its origin and theoretical description are still under debate. In this work, as one of the simplest model for describing vibrational properties in disordered systems, inhomogeneous elastic wave equation, is solved in real space without using perturbative approach as previous works. In real space solution, the BP associated flat dispersion relation can be obtained, localized vibration in exponential decay in soft spot can be observed, and the fluctuation length of shear modulus dependent BP frequency is also confirmed. These features have been reported in recent progresses but missed within perturbative approach. This work unify divergent and controversial conclusions of BP within a simple model of fluctuating shear modulus under clear visualization.
\end{abstract}
\maketitle

\color{blue}\textit{Introduction} \color{black} -- 
Vibration of amorphous materials have been extensively observed with excess vibrational density of states (VDOS) than crystalline counterparts in the low frequency regime known as Boson peak (BP).\cite{PhysRevB.4.2029,Hu2022,PhysRevResearch.5.023055,Ahart2017} For crystalline materials, Debye's theory with consideration of volume taken in wave vector space of each vibrational mode predicts that VDOS should obey power law in function of frequency $g(\omega) \propto \omega^{d-1}$ up to a cut off wave vector known as Debye's wave vector $k_D$.\cite{thebook} Here $d$ is the dimension of material. Debye's theory have two assumptions that invalidate it for amorphous materials. The two assumptions are S1. the material need to be homogeneous hence volume taken in wave vector space is same for all allowed modes, and S2. the dispersion relation need to be linear $\omega = c k$ where $c$ is the speed of sound. 

During debate with decades of research on the origin of BP, there are three features of the vibration anomaly have been concluded for ordinary amorphous solids, C1. the anomaly vibration is contributed by transverse wave,\cite{Shintani2008,Ren2021,PhysRevLett.106.225501} C2. the anomaly vibration is (quasi) localized,\cite{Hu2022,Zhang2011,PhysRevResearch.5.023055,PhysRevLett.133.188302,10.21468/SciPostPhys.15.2.069,PhysRevLett.127.215504} and C3. there is a flat dispersion relation $\omega = \omega_{BP}$ appearing at the frequency of BP $\omega_{BP}$.\cite{PhysRevLett.133.188302,Hu2022,PhysRevResearch.5.023055,PhysRevB.98.174207} In addition, there are two features that are not universally verified yet but still indicated by serval researches, they are, C4. The frequency of BP is mainly determined by the size of localized vibration,\cite{10.1063/5.0210057,Jiang_2024,PhysRevLett.133.188302,PhysRevB.101.174311} and C5. localized vibration appear with soft spots.\cite{Hu2022,PhysRevB.43.5039,PhysRevB.76.064206}

Conclusions C2, C4 and C5 indicate that vibration and dynamical properties in amorphous solids are not homogeneous, which against assumption S1 of homogeneity in Debye's theory. While conclusion C3 indicate presence of flat dispersion relation which invalidate assumption S2 of linear dispersion relation. Various theories have been proposed in the past long journey to explain the origin of BP, including, soft mode vibrations,\cite{PhysRevB.43.5039,PhysRevB.76.064206}, anharmonic effects,\cite{PhysRevLett.122.145501,PhysRevB.67.094203} quasi-localized vibrations,\cite{PhysRevB.67.094203,PhysRevB.43.5039} and inhomogeneous elasticity in perturbative field-theoretical scheme.\cite{PhysRevLett.98.025501,PhysRevLett.100.137402} However, the well observed features in experiments and simulations from C2 to C5 are never unified in a single theoretical framework because theories are usually constructed in Fourier space while experiments and simulations are done in real space. 

In this work, the author solve inhomogeneous elastic wave equation in real space with relaxed random initial vector field in 2D. Then the current correlation function analysis is applied to obtain dispersion relations of linear longitudinal, linear transverse and flat transverse branches from the real space solution. The exponential decaying vibration at soft spots is also confirmed in real space solution. Additionally, the frequency of flat dispersion relation is confirmed determined only by the scale of shear modulus fluctuation. The results include all the five features from C1 to C5 observed in experiments and simulations, and all the results start from an elastic wave equation with fluctuating shear modulus. This work suggest that inhomogeneous elasticity can be considered as a good picture to interpret the physics of low frequency vibrational anomaly in amorphous solids.

\color{blue}\textit{Model and analysis} \color{black} -- 
In the theoretical framework, the structure of amorphous solid is assumed to induce inhomogeneous elasticity. The vibration properties can be governed by an elastic wave equation with inhomogeneous elasticity,\cite{PhysRevLett.98.025501,thebook}
\begin{equation}
    \rho \dfrac{\partial^2}{\partial t^2} \boldsymbol{u} - \nabla (M \nabla \cdot \boldsymbol{u}) + \nabla \times (G \nabla \times \boldsymbol{u}) = 0, \label{waveeq}
\end{equation}
where $\boldsymbol{u}(\boldsymbol{r},t)$ is displacement field in function of position $\boldsymbol{r}$ and time $t$, $\rho$ is density, $G(\boldsymbol{r})$ is shear modulus with spatial fluctuations, and $M(\boldsymbol{r}) = K + G(\boldsymbol{r})$ is longitudinal modulus with $K$ the bulk modulus assumed without spatial fluctuations. The speed of sound are therefore $c_{L(T)} = \sqrt{M(G)/\rho}$. Eq.\eqref{waveeq} will be solved in 2D for computational cost consideration and better visualization.

Different from perturbative approach in,\cite{PhysRevLett.98.025501,PhysRevLett.100.137402} real space solution need boundary condition and initial condition to solve Eq.\eqref{waveeq} in real space. In this work, open boundary is used which means elastic wave will be reflected when reach the boundary. The initial condition $\boldsymbol{u}(\boldsymbol{r},t)_{t=0}$ can use random vector field $\boldsymbol{R}$ generated by random number generator after some physical setup. Here the randomness is used as thermal motion of each microscopic volume elements in the system, hence there should be a maximum wave vector, Debye wave vector $k_D$, above which thermal excitations in wave vector space are not allowed. The second setup is that soft area should have larger mean displacement than hard area due to elastic force balance according to Hooke's law. With these two consideration, the initial condition can be chosen as,
\begin{equation}
    \boldsymbol{u}(\boldsymbol{r},t)_{t=0} = f_{k \leq k_D}[\boldsymbol{R}_L(\boldsymbol{r})] \dfrac{\Bar{M}}{M(\boldsymbol{r})} + f_{k \leq k_D}[\boldsymbol{R}_T(\boldsymbol{r})] \dfrac{\Bar{G}}{G(\boldsymbol{r})}. \label{thermalization}
\end{equation}
Here, $\boldsymbol{R}_L + \boldsymbol{R}_T = \boldsymbol{R}$ are the longitudinal (L) and transverse (T) components of random vector field. $f_{k \leq d_D}[]$ works as a filter that survive only the Fourier components with wave vector smaller than $k_D$. The factors $\Bar{M}/M(\boldsymbol{r})$ and $\Bar{G}/G(\boldsymbol{r})$ are used to avoid force imbalance induced by fluctuation of shear modulus as the second setup, where $\Bar{M}(\Bar{G})$ denote their average value.

The spatial fluctuations of shear modulus, it can be set using scalar random field $R$, as,
\begin{equation}
    G(\boldsymbol{r}) = \alpha f_{k = k_G}[R],\label{distribution}
\end{equation}
where $k_G = 1/ l_G$ set the length scale of shear modulus fluctuations, $\alpha \in [0, \alpha_c]$ set the fluctuation strength. Here $\alpha_c$ is critical fluctuation when $G(\boldsymbol{r} = 0)$ appear locally. The length scale of modulus fluctuations should be larger than the shortest wave length, that is $k_G < k_D$.

The real space solution of displacement field $\boldsymbol{u}(\boldsymbol{r},t)$ can be obtained under controlling of three parameters, Debye wave vector $k_D$, fluctuation length of shear modulus $l_G$ and fluctuation strength of shear modulus $\alpha$. With the real space solution, current correlation function in frequency domain can be obtained according to,\cite{ccf}
\begin{equation}
    C_{L(T)}(\boldsymbol{k}, \omega) = \mathcal{F}_t [\langle \boldsymbol{v}_{L(T)}(\boldsymbol{k}, t) \cdot \boldsymbol{v}_{L(T)}(-\boldsymbol{k}, 0) \rangle],
\end{equation}
where $\mathcal{F}_t[]$ denotes the Fourier transform on time domain, $\boldsymbol{v}_{L(T)} (\boldsymbol{k}, t)= \mathcal{F}_{\boldsymbol{r}}[\boldsymbol{v}_{L(T)}(\boldsymbol{r}, t)]$ is the Fourier transform in spatial domain of velocity field which is given by $\boldsymbol{v}_{L(T)}(\boldsymbol{r}, t) = \partial_t \boldsymbol{u}_{L(T)}(\boldsymbol{r}, t)$, and $\langle \rangle$ denotes ensemble average. The VDOS can be obtained as $g_{L(T)}(\omega) = \int C_{L(T)}(\boldsymbol{k}, \omega) / \Bar{v}^2 d\boldsymbol{k}$.

To see the vibration under BP frequency from real space solution, one can use a filter to remove the components of other frequency, as,
\begin{equation}
    \boldsymbol{v}_{L(T),\omega \approx \omega_{BP}} = f_{\omega \approx \omega_{BP}}[\boldsymbol{v}_{L(T)}],\label{localmotion}
\end{equation}
where $\omega_{BP}$ is the frequency of BP and also flat dispersion relation.

\color{blue}\textit{Results and discussion} \color{black} -- 
Before presenting the results, it needs to clarify that all units involved in this work are used self consistently, like Leonard Jones units in simulations,\cite{Hu2022} hence all quantities will be presented just in number without unit behind. 

\begin{figure}[ht!]
    \centering
    \includegraphics[width=\linewidth]{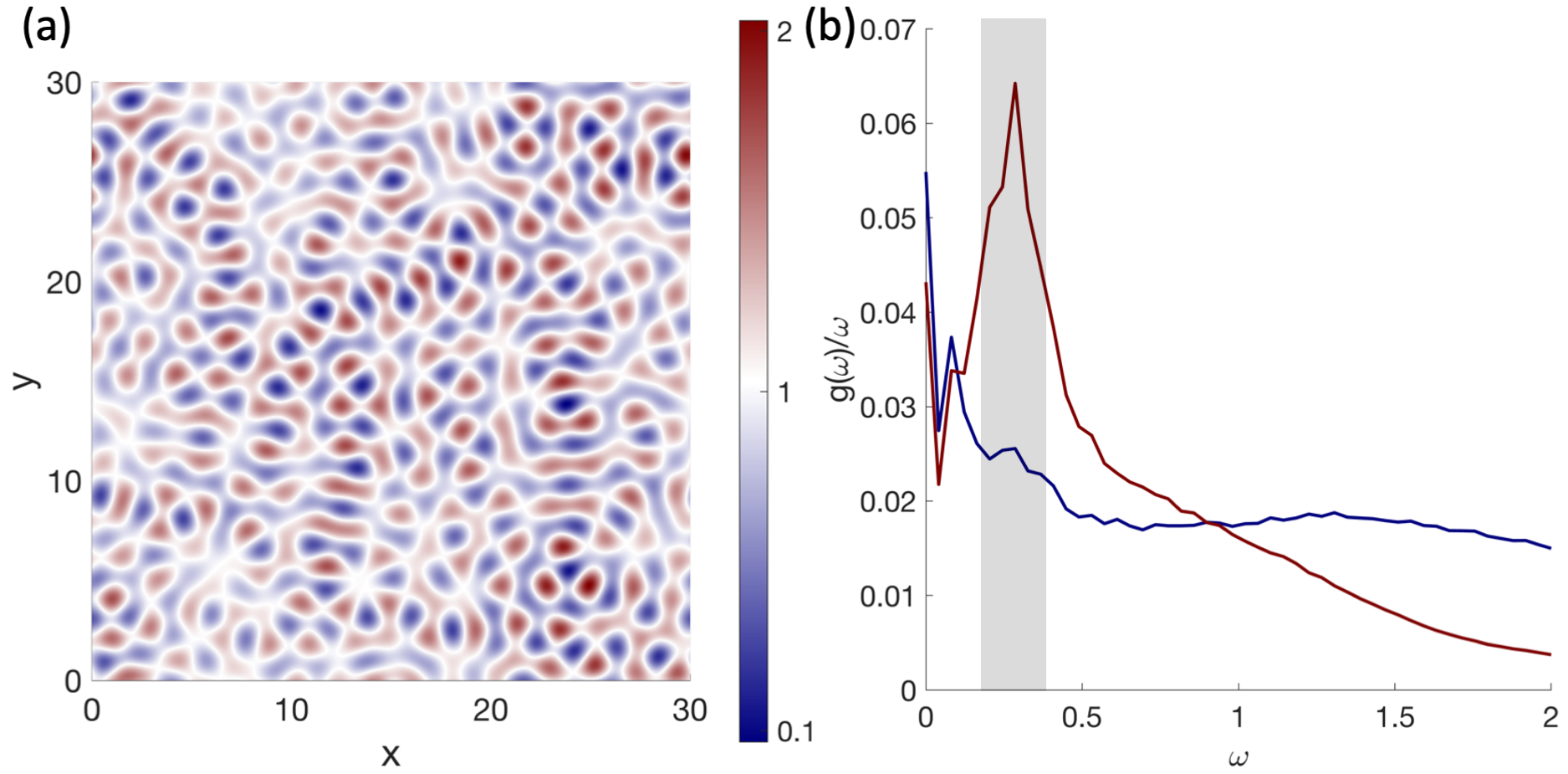}
    \caption{\textbf{inhomogeneous speed of sound $c_T (\boldsymbol{r})$ and VDOS}. \textbf{(a)}, the real space distribution of inhomogeneous speed of sound $c_T (\boldsymbol{r})$. Color from blue to red in panel (a) indicate increasing $c_T$ as shown by the color bar. The average transverse speed of sound $\bar{c_T} = 1$.  \textbf{(d)}, the reduced VDOS. Blue and red in panel (b) are used to distinguish longitudinal and transverse respectively. The gray bar show the position of flat dispersion relation and BP. The parameters involved to obtain these results are, $K = 1.7$, $\bar{G} = 1$, $\alpha = 2$, $\rho = 1$, $k_D = 1.5$, $k_G = 0.5$, the size of simulated area is $30\times 30$, and maximum observation time is $50$.}
    \label{figdos}
\end{figure}

Fig.\ref{figdos}-(a) show the distribution of speed of sound $c_T (\boldsymbol{r}) = \sqrt{G(\boldsymbol{r}/\rho)}$ in real space. The distribution have only one wave vector $k_G$ survived in Fourier space and therefore it have only one length scale $l_G = 1/ k_G$. For the distribution of different length scales, please see Fig.\ref{figkg}. The distribution of shear modulus is not well confirmed in experiments or simulations yet due to difficulties on the definition of local elasticity. However, Eq.\eqref{distribution} provide a ideal and clean distribution with only one length scale, that can be a good theoretical consideration. We will see in later, the single length scale of distribution controls frequency of flat dispersion relation in ideally manner, the frequency is inverse to the length scale.

BP regime of VDOS is highlighted by gray color in Fig.\ref{figdos}-(b). It can be seen that BP appear only for transverse part. The data points at very low frequency close to zero are not good defined since it needs super long observation time to take out near zero frequency data. In this work, the maximum observation time is $50$, it corresponding to the frequency $1/50$ low enough for studying BP at frequency around $0.24$.

\begin{figure}[ht!]
    \centering
    \includegraphics[width=\linewidth]{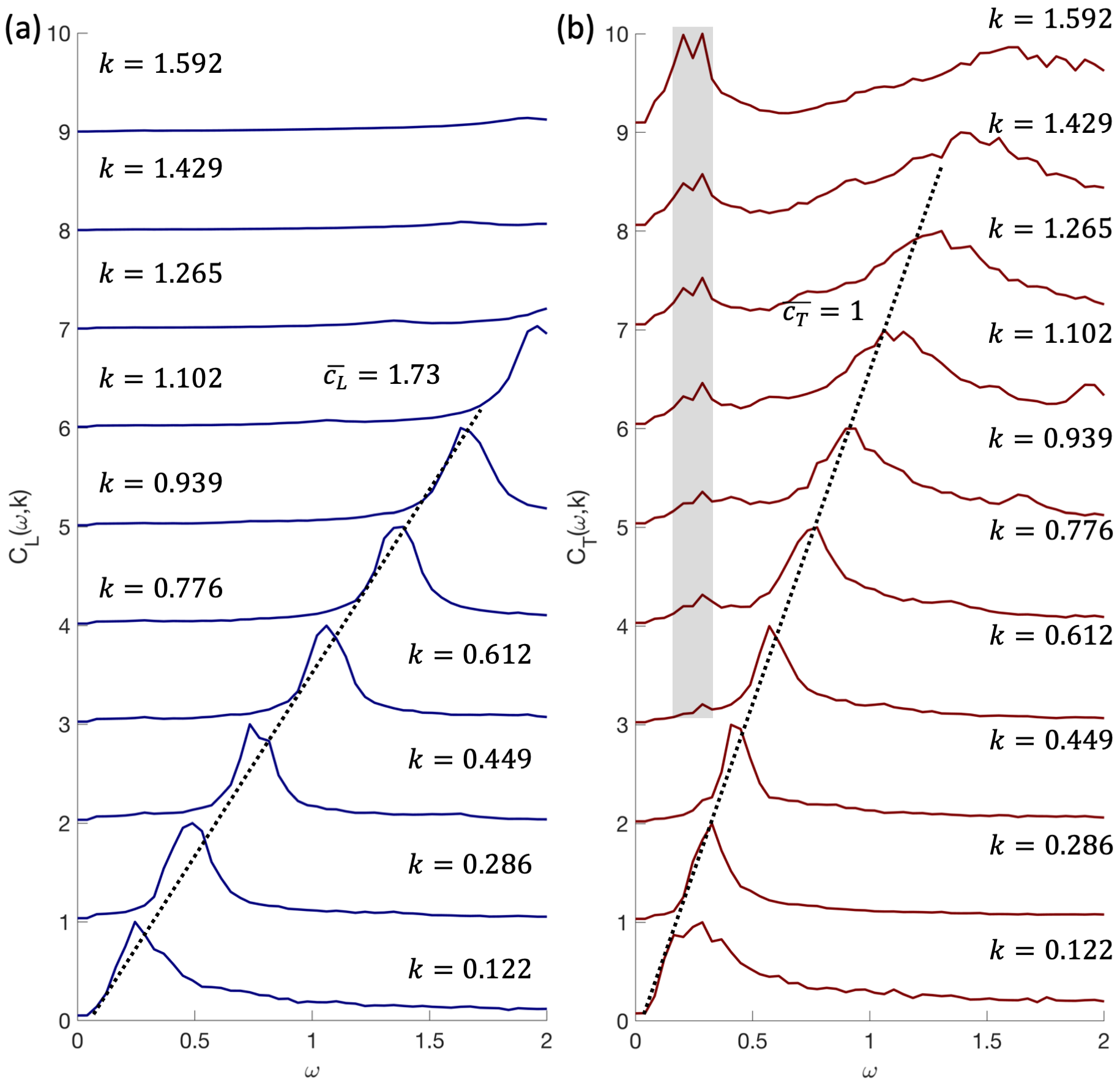}
    \caption{\textbf{Current correlation function}. The longitudinal \textbf{(a)} and transverse \textbf{(b)} current correlation function at different wave vectors. The dashed lines show their linear dispersion relations with speed of sound $\Bar{c_L} = 1.73$ and $\Bar{c_T} = 1$ respectively. The gray bar show the additional flat dispersion relation.}
    \label{figdsf}
\end{figure}

For comparing dynamical properties in different magnitude of wave vector $k = |\boldsymbol{k}|$, current correlation functions $C_{L(T)}(k,\omega)$ are shown in function of $\omega$ with different $k$ in Fig.\ref{figdsf}. Here, the vector $\boldsymbol{k}$ have been averaged angularly to obtained magnitude result $k$. In Fig.\ref{figdsf}, longitudinal current correlation function indicate only a linear dispersion relation with speed of sound $\bar{c_L} = 1.58$ while longitudinal current correlation function indicate not only a linear dispersion relation with speed of sound $\bar{c_T} = 1$ but also a flat dispersion relation $\omega = 0.24$ independent of $k$. The presence of flat dispersion relation should be responsible for BP in reduced VDOS as observed in references for conclusion C3.\cite{PhysRevLett.133.188302,Hu2022,PhysRevResearch.5.023055,PhysRevB.98.174207} There are two minor features need to be noticed in Fig.\ref{figdsf}. The first feature is that linewidth of transverse linear branch can be greatly increased when $k$ increase. This effect can be well understood as wave with large $k$ suffering stronger scattering due to presence of disorder. The second feature is the peak of flat dispersion relation on current correlation function $C_{T}(k,\omega)$ is weak when $k$ is small to reach linear dispersion relation. This feature is also observed in Ref.\cite{Hu2022} where the authors attribute it to interaction between phonon and localized mode.

\begin{figure}[ht!]
    \centering
    \includegraphics[width=\linewidth]{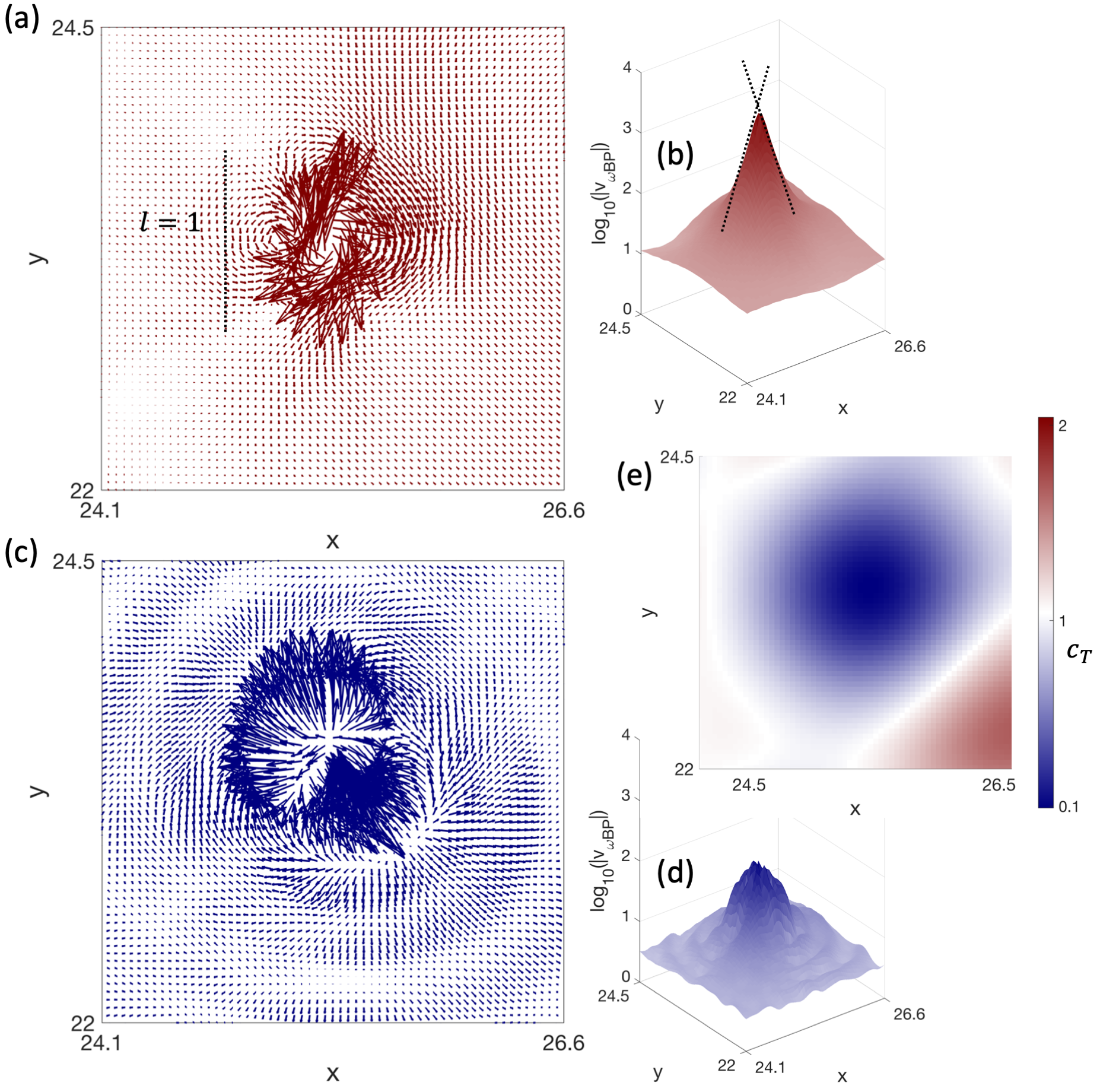}
    \caption{\textbf{Motion under BP frequency}. \textbf{(a)}: A snapshot of transverse velocity field under BP frequency, $\boldsymbol{v}_{T,\omega \approx \omega_{BP}}$. \textbf{(b)}: The absolute value of panel (a) in linear-log plot. The transverse part is exponentially decay and therefore to be localized. \textbf{(c)}: A snapshot of longitudinal velocity field under BP frequency at same position, $\boldsymbol{v}_{L,\omega \approx \omega_{BP}}$. \textbf{(d)}: The absolute value of panel (c) in linear-log plot. The longitudinal part is not exponentially decay and hence is not localized. \textbf{(e)}: Show the fluctuation of transverse speed of sound $c_T (\boldsymbol{r})$ due to fluctuation of shear modulus at same area. The localized motion appears at the softened valley. The color blue and red are used to distinguish longitudinal and transverse parts for panels (a), (b), (c) and (d). The color blue and red are used show the distribution in panel (e) as shown in the color bar. The size of localized motion is is approximately $l = 1$, which is close to $l = 1/(2 k_G)$ with $k_G = 0.5$ as an input parameter.}
    \label{figlocal}
\end{figure}

As mentioned before, the flat dispersion relation is considered as a result of localized vibration,\cite{Hu2022,PhysRevLett.133.188302,PhysRevB.67.094203} which is different from acoustic wave described by linear dispersion relation. One of the Advantage of real space solution is accessible visualization of such localized vibration. By using Eq.\eqref{localmotion}, vibration under frequency of flat dispersion relation can be observed. Fig.\ref{figlocal} show a snapshot of such vibration around that frequency with separated longitudinal and transverse components. It can be seen that the vibrational intensity $|\boldsymbol{v}_{\omega \approx \omega_{BP}}(\boldsymbol{r})|$ is stronger in the soft spot than other regime for both longitudinal and transverse components. However, the decay rate is exponential for transverse vibration while longitudinal vibration is not. Exponential decay is usually a signature of localization in various condensed matter phenomena.\cite{Segev2013} Therefore the exponential decay of transverse vibration intensity indicate they are localized. By comparing Fig.\ref{figdsf} and Fig.\ref{figlocal}, it can be confirmed that localization and flat dispersion relation appear together in transverse wave. Although vibrational intensity is also stronger at the soft spot for longitudinal components, but there is neither exponential decay nor additional flat dispersion relation. 

In real space, the size of localized vibration can be easily estimated. As an example, in Fig.\ref{figlocal}-(a), the size of localized vibration is around $l=1$ which is half of shear modulus fluctuating length scale $l_G$, $l = l_G/2 = 1/(2k_G)$. The size of localized vibration is roughly also the size of soft spot shown in Fig.\ref{figlocal}-(e). Here the relation of length scale can be interpreted in following way. $l_G = 1/k_G$ set the length scale of shear modulus fluctuation, similar to sine function $\mathrm{sin}x$ with wavelength $2\pi$. However the soft spot is only where $\mathrm{sin}x < 0$, hence the soft spot is only $\pi$ only a half of wavelength $2\pi$. The situation is similar in Fig.\ref{figlocal}-(e), size of soft spot is basically half of $l_G = 2$. And the size of soft spot determine the size of localized vibration. 

\begin{figure}[ht!]
    \centering
    \includegraphics[width=0.6\linewidth]{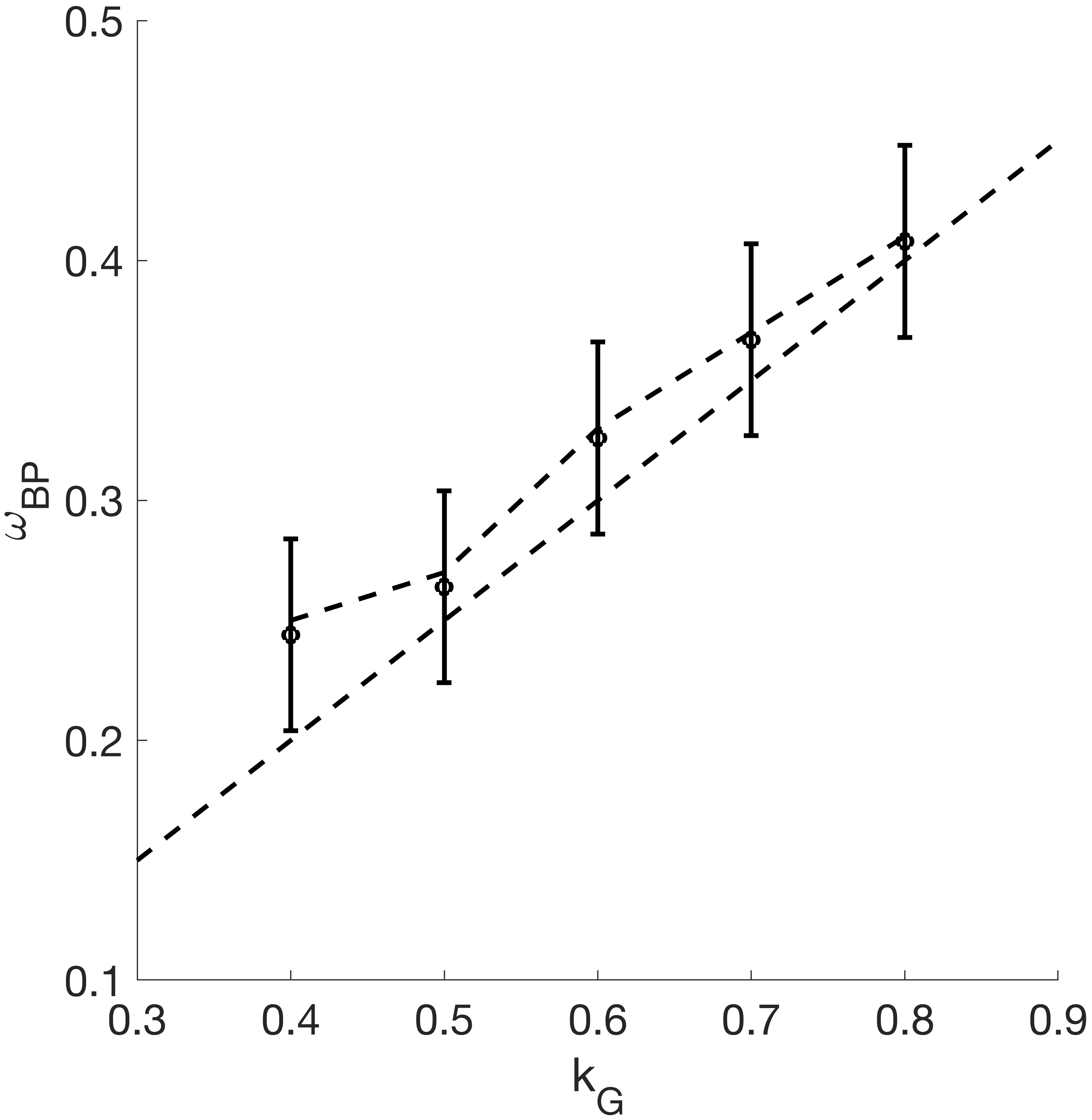}
    \caption{\textbf{Frequency of BP}. The frequency of BP, $\omega_{BP}$, in function of fluctuation length scale. There is roughly $\omega_{BP} = \Bar{c_T} k_G /2$.}
    \label{figkgandwbp}
\end{figure}

In previous researches, the frequency of localized vibration is assumed to be determined by its size,\cite{PhysRevLett.133.188302,Jiang_2024,PhysRevB.101.174311} $\omega_{local} = \pi c/l_{local}$ with $c$ the average speed of sound. With the size and distribution of localized mode obtained from simulation, the formula $\omega_{local} = \pi c/l_{local}$ is able to estimate frequency of BP quantitatively. Accroding to previous discussion, the size of localized vibration in present model $l=1/(2k_G)$ is easy to control by input parameter $k_G$, and the size distribution is ideally Delta function $\delta_{l,1/(2k_G)}$ due to Eq.\eqref{distribution}. Therefore, the frequency of BP can be drove to be $\omega_{BP,theory} = 2 \bar{c_T} k_G$. Fig.\ref{figkgandwbp} show the frequency of BP in function of various fluctuation length inverse $k_G$, the function can be written in form of $\omega_{BP} = \bar{c_T} k_G / 2$. Except a factor of $4$, real space solution of Eq.\eqref{waveeq} can explain BP to be length scale of shear modulus fluctuation dependent. 

To this stage, all five features of BP from C1 to C5 mentioned previously have been observed in real space solution of inhomogeneous elastic wave equation Eq.\eqref{waveeq}. Although the inhomogeneous elasticity model was considered to underestimates the importance of localized vibration,\cite{PhysRevLett.130.236101,PhysRevLett.123.055501} here, in this work, it have been shown the localized vibration should be responsible for presence of flat dispersion relation and hence BP in also inhomogeneous elasticity model. However, it still need to be careful when compare inhomogeneous elasticity model with real systems due to definition problem of elasticity in particle level scale.

\color{blue}\textit{Conclusions}\color{black} -- 
In summary, the author report low frequency dynamical anomaly in both Fourier and real space based on real space solution of inhomogeneous elastic wave equation. It is shown that real space solution can give excess VDOS, BP, at low frequency by using current correlation function analysis. By checking current correlation function, it is confirmed that BP is induced by a flat dispersion relation appeared in addition to acoustic linear branch for transverse components. In real space, the exponentially decaying localized vibration is also observed to appearing in soft spot of fluctuating shear modulus. The size of localized vibration is shown to be identical to the fluctuating length scale of elasticity. And the frequency of localized vibration is also confirmed inversely depending on the fluctuating length scale of elasticity via $\omega_{BP} = c_{T} / (2 l_G)$. These conclusions are agree with recent progresses of BP on experiments and simulations, which suggest that inhomogeneous elasticity can be accepted as a qualitative theoretical model to unifying BP, flat dispersion relation and low frequency localized vibration those universal phenomena of amorphous solids in a simple framework.

\color{blue}{\it Acknowledgments} \color{black} --  The author would like to thank Matteo Baggioli, Hajime Tanaka, Massimo Pica Ciamarra, Yinqiao Wang, Qing Xi and Shivam Mahajan for very illuminating discussions on BP. The author also would like to thank Tengzhou Zhang for his instructive suggestions on computational method. The author acknowledge the support of the Shanghai Municipal Science and Technology Major Project (Grant No. 2019SHZDZX01).

\bibliographystyle{apsrev4-1}


\newpage
\onecolumngrid
\appendix 
\clearpage
\renewcommand\thefigure{S\arabic{figure}}    
\setcounter{figure}{0} 
\renewcommand{\theequation}{S\arabic{equation}}
\setcounter{equation}{0}
\renewcommand{\thesubsection}{SM\arabic{subsection}}
\section*{\Large Supplementary Information}
\textbf{In this Supplementary Information (SI), details on the methods used and further analysis about this work will be provided.}

\subsection{Distribution of speed of sound in different length scale}
\begin{figure}[ht!]
    \centering
    \includegraphics[width=0.8\linewidth]{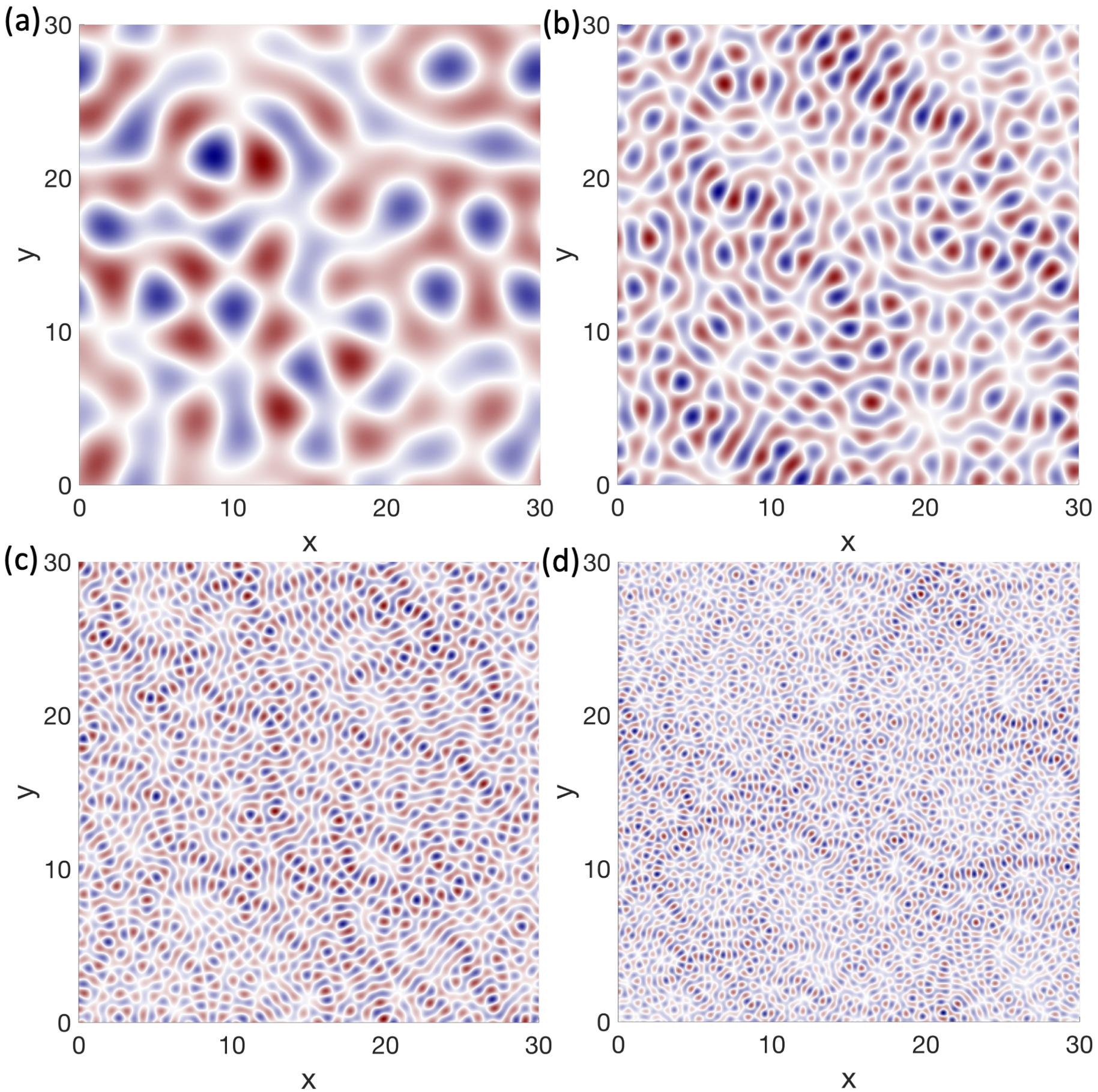}
    \caption{\textbf{Distribution of speed of sound in different length scale}. The distribution of $c_T(\boldsymbol{r})$, from \textbf{(a)} to \textbf{(d)}, the parameters are $k_G = 0.2$, $k_G = 0.5$, $k_G = 1$ and $k_G = 1.5$. The color from blue to red indicate stronger value.}
    \label{figkg}
\end{figure}

Fig.\ref{figkg} show the distribution of speed of sound $c_T(\boldsymbol{r})$ obtained by,
\begin{equation}
    G(\boldsymbol{r}) = \alpha f_{k = k_G}[R]
\end{equation}
with $c_T = \sqrt{G/\rho}$ for different parameters $k_G$. 

\subsection*{Current correlation function in color}\label{meth}
\begin{figure}[ht!]
    \centering
    \includegraphics[width=0.8\linewidth]{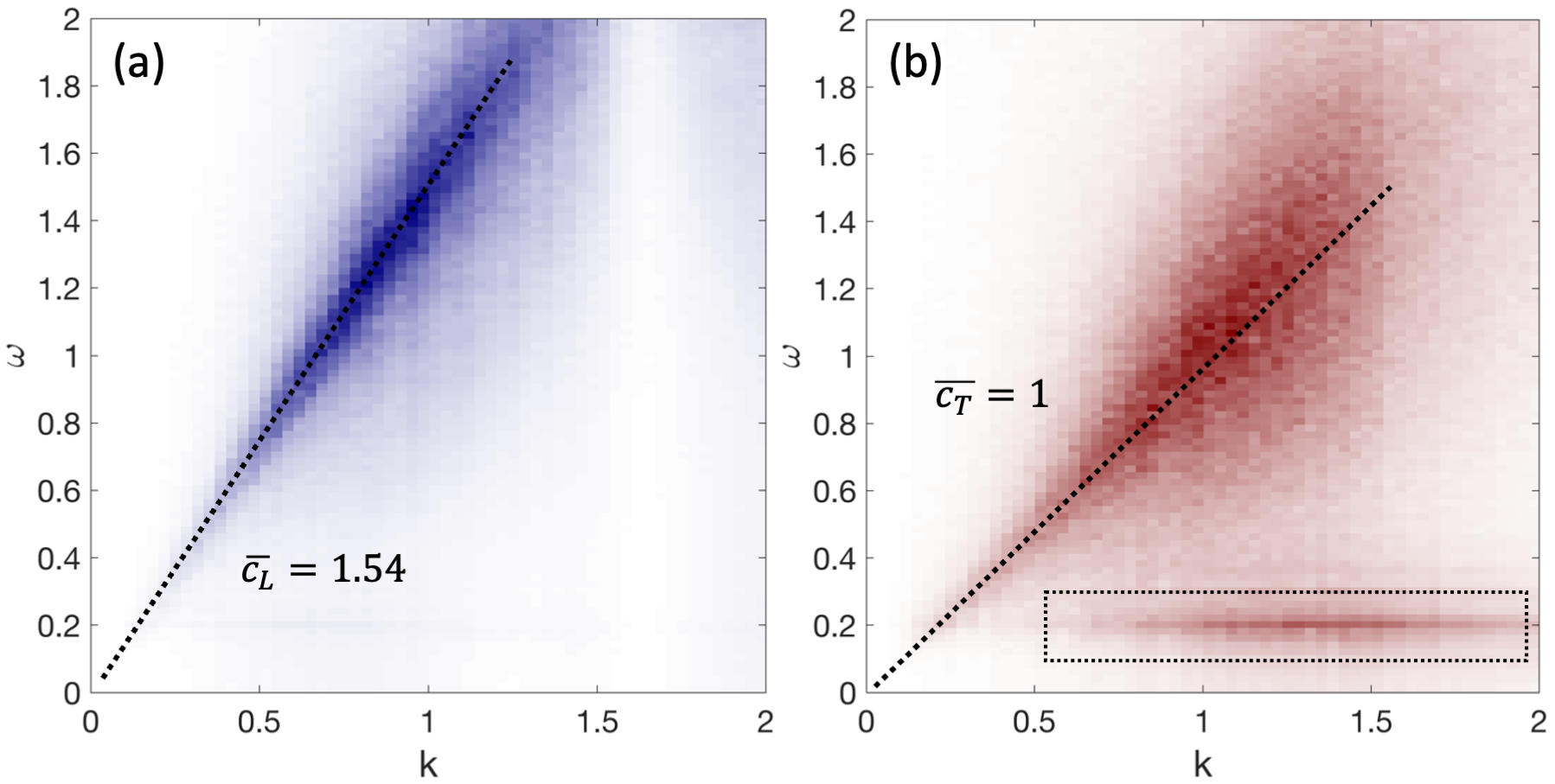}
    \caption{\textbf{Dispersion relation}. \textbf{(a)}, the longitudinal current correlation function. \textbf{(b)}, the transverse current correlation function. Black dashed line show their linear dispersion relation with average speed of sound $\Bar{c_L} = 1.54$ and $\Bar{c_T} = 1$. The black dashed bar show the additional flat dispersion relation. The darker the stronger intensity.}
    \label{figdsdcolor}
\end{figure}

Fig.\ref{figdsdcolor} show the current correlation function $C_{L(T)}(\omega, k)$ in color plot. One can see the linear dispersion relations for longitudinal and transverse acoustic branches. And the additional flat dispersion relation.

\end{document}